# On the Implementation of a deterministic secure coding protocol using polarization entangled photons

Martin Ostermeyer, Nino Walenta

*University of Potsdam, Institute of Physics and Interdisciplinary Center for Photonics Potsdam*
*Am Neuen Palais 10, 14469 Potsdam, Germany*

**Abstract:** We demonstrate a prototype-implementation of deterministic information encoding for quantum key distribution (QKD) following the ping-pong coding protocol [1]. Due to the deterministic nature of this protocol the need for post-processing the key is distinctly reduced compared to non-deterministic protocols. In the course of our implementation we analyze the practicability of the protocol and discuss some security aspects of information transfer in such a deterministic scheme.

# 1 INTRODUCTION

The quantum key distribution protocol invented by Bennett and Brassard (BB84) [2] has become a standard in quantum key distribution (QKD). In practical implementations of this protocol single photons are used as qubit carriers and the information transferred between the sender Alice and the receiver Bob is encoded in statistically swapped basis sets, e.g. of the photon polarization. QKD protocols based on entangled qubits have the potential of improved "security features" compared to schemes based on single qubits as it was stated in [3], though theoretical and experimental analysis needs to be carried out to support this view.

QKD using entangled photons has been demonstrated in a range of experiments before. Polarization entangled pairs of photons were used as an approximation of a conditional single photon source in the single qubit BB84 protocol [5]. They were also applied to protocols like the Ekert protocol [6] where both photons are used for the key transmission [5] or the "six state protocol" [7] implemented by Enzer et.al [8]. Another class of experiments uses phase encoding [9], energy-time entanglement [10] or the SARG protocol [11] based on time bin qubits instead of polarization entanglement. QKD via fibres was recently demonstrated beyond 100 km distance with qubit-error rates (QBER) of 8.9 % [12] and 5 % [13] respectively. An overview of methods and techniques in QKD can be found in [14] and [31].

In this paper we discuss our experimental implementation of the encoding procedure of a specific example of deterministic QKD-protocols [1]. Although the inventors of the protocol also claim the capability of the protocol for quasi-secure-direct communication, they admit that the direct secure communication variant of their proposal does not allow for "perfect secure communication" [1]. Thus, we concentrate on the QKD discussion, only.

In contrast to other concepts of QKD with entangled photons the ping-pong coding protocol belongs to the class of deterministic protocols. There is expected to be less need for classical post-processing of the key given that in the ideal lossless case, the scheme operates in a deterministic fashion. The deterministic character of the protocol provides the opportunity for using all transmitted photons for key generation where no bits have to be discarded. This is a distinct difference compared to other protocols which are based on the BB84 or Ekert-scheme and it is the base for its potential for high key transfer rates. Another advantage of the deterministic characteristic is the higher eavesdropping detection probability compared to e.g. the BB84 protocol [1]. The drawbacks of the deterministic characteristic within practical implementations of the protocol are discussed in section 4.

According to the protocol, one bit of information gets encoded in the relative phase of the entangled qubits. Both qubits are generated and detected at Bob's site, but only one qubit, the travel qubit, is sent to Alice (ping) and then back to Bob (pong). The other qubit, the home qubit, is stored at Bob's site. From the travel qubit alone the encoded information cannot be extracted. Indeed, decoding is possible only if a Bell state measurement is performed on both qubits together to evaluate the state by means of the correlations between them. With the generic ping-pong characteristic of the protocol the problem of entanglement distribution is transferred to the storage of the home photon at Bob's device for the duration of the travel photon propagation.

The ping-pong coding protocol inspired a number of authors to further develop the idea of ping-pong-protocols (e.g. [15, 16]). Different aspects of the security of the original protocol were discussed in the literature. A serious comment on the application of the protocol on lossy quantum channels is presented by Wojcik in [17]. He pointed out how an eavesdropper (Eve) could hide undiscovered in the loss of the quantum channel if the transmission efficiency is lower than 60 %. As proposed in the same publication, however, Eve can be discovered by an additional test during the control run (see section 2). In [18] Cai confirms the security of the ping-pong-protocol but discusses a possible denial-of-service (DoS) attack. Such an attack could be detected by post-processing the detection events which, however, would undermine the deterministic character of the protocol. Zhang addresses further attacks in [19, 20]. But these attacks seem to have flaws which have been addressed recently by the inventors of the ping-pong-protocol [21]. However, a full security proof of the ping-pong-protocol in case of noisy channels was not given yet. To prove security for the modified variant of the protocol within our implementation would be beyond the scope of this paper and remains to be given in the future. Still, the deterministic character of the protocol attracted our attention and resulted in the here presented prototype implementation. Individual security aspects of our implementation are discussed in section 4.

## 2   Implementation of ping-pong encoding using polarization entangled photons based on a dense coding scheme

In this section we will present an exemplary implementation of the encoding part of the ping-pong-protocol. This will enable us to discuss key aspects in secure deterministic encoding on practical grounds in section 4. To suit the needs for this specific protocol we altered a dense coding scheme with polarization entangled photons [27] as will be pointed out below. To benefit mostly from the deterministic character of the encoding procedure a deterministic source of entangled photon pairs would be appropriate. Since those sources are still objects of research (see e.g. [28]) and hence not available off the shelf we have exemplarily chosen to set up an entangled photon source based on parametric fluorescence. The ramifications of this decision are discussed in section 4.

To encode data in a binary alphabet two of the four Bell states are sufficient. Following the original paper we use the $|\psi^+\rangle$ state as 0 and $|\psi^-\rangle$ state as 1 with

$$|\psi^\pm\rangle = \frac{1}{\sqrt{2}}\left[|H\rangle_h|V\rangle_t \pm |V\rangle_h|H\rangle_t\right].$$

Here, index h or t denotes home or travel photon analogous to the home and travel qubit described in the original protocol. To switch between $|\psi^+\rangle$ and $|\psi^-\rangle$, a phase shift of π between horizontal (H) and vertical (V) polarization has to be applied.

Our implementation of the ping-pong-protocol is depicted in Fig. 1. The polarization entangled photons are generated by type II spontaneous parametric down-conversion (SPDC) [23] in a beta-Barium-Borate crystal of 1 mm length. For this proof of principle experiment the BBO crystal is pumped by a mode locked Picosecond laser with pulse duration of 8 ps, average power of 300 mW, and repetition rates of 85 MHz at 355 nm wavelength. The pump spot in the BBO-crystal has a diameter of 180 µm. The photons are detected with fibre coupled avalanche diode detectors with active quenching (AQR-14 by Perkin Elmer). Their specified quantum efficiency at 710 nm is 72%.

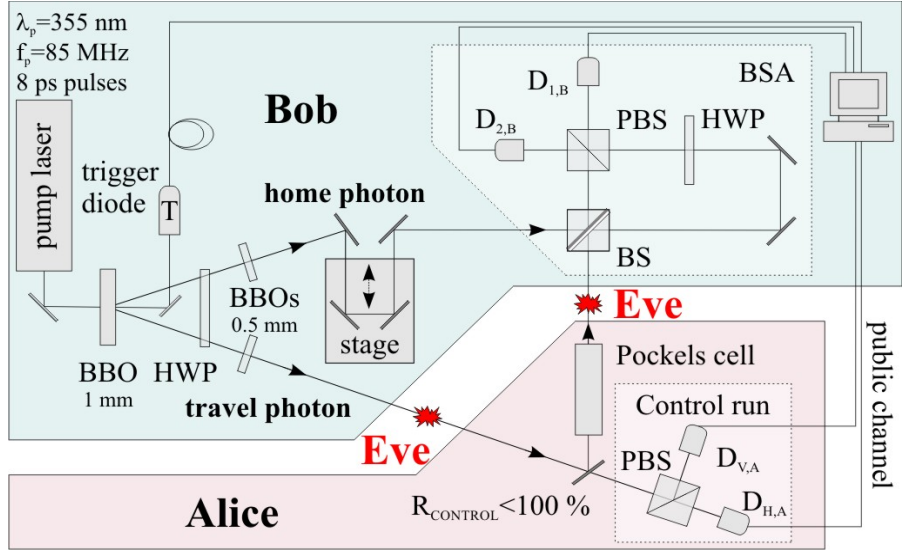

Fig. 1: Set up for the optical implementation of the ping-pong-coding protocol. PBS denotes polarizing beam splitters, BSA the district for the Bell state analysis, HWP a half wave plate, and BBO a beta-Barium-Borate crystal. The translation stage at Bob is only active for the test of our Bell state analysis (see text).

The beam propagation of the intersection area of the two SPDC emission cones into the single mode fibers of the photon detectors is designed according to the mode matching procedure described in [24]. The resulting total bandwidth that is coupled into the fibers corresponds to 28 nm (FWHM). The counting rates $R_1$ and $R_2$ of the detected single photons generated by SPDC in the setup in Fig. 1 is $1.7 \cdot 10^5$ counts/second in each of the detectors $D_{1,B}$ and $D_{2,B}$ at Bob. The coincidence rate $R_c$ of the detected photon pairs amounts to 17 % of this value and is called correlation coefficient. The correlation coefficient can be defined by the $C = R_c/(R_1 R_2)^{0.5}$ with $R_{1,2}$ denoting the single photon counting rates at detectors $D_{1,B}$ and $D_{2,B}$ respectively and $R_c$ being the coincidence counting rate of the two detectors. Since the correlation coefficient is limited by the total detection efficiencies $\eta_{1,2}$ for the detection at $D_{1,B}$ and $D_{2,B}$ it holds $C = \eta_1 \cdot \eta_2$. This total detection efficiency includes all losses e.g. fiber coupling losses and can be estimated to be $\eta_{1,2}$ = 41 % ($\sqrt{0.17} = 0.41$).

The ratio between the two pair and single pair generation rate can be calculated from the single photon counting rate and the detection efficiencies according to [22]. It results to around $4 \cdot 10^{-2}$. Behind the BBO-crystal the SPDC emission is corrected for runtime differences by a halfwave plate and one compensation crystal each per intersection area of the SPDC emission cones. The polarization entanglement of the source is characterized in the setup using Bob's detectors with an S parameter of 2.51±0.0049. The visibility of the produced entangled state $|\psi^+\rangle$ is better than 93 %.

The home photon which stays with Bob can be stored in a delay line e.g. a fibre loop. Within this proof of principle experiment in our lab Alice and Bob are just separated by about one meter. Thus, we use multiple reflections between mirrors to store the home photon. The maximum length of the fiber loop in an implementation using polarization entanglement is given by the wavelength dependent attenuation and depolarization mainly due to birefringence in the fiber. For longer distance implementations one would change the wavelength and work with photons within the low attenuation telecom band at 1550nm. Recently Hübel at al. [25] showed that using a non zero dispersion shifted fiber for 1550 nm and a compensation scheme for the polarization drift distances of 100 km are feasible for the fiber transmission of polarization encoded qubits.

The travel photon is sent to Alice (ping) where it is randomly reflected by a partially reflecting mirror. Thus, with a given probability $R_{control}$ the protocol is set to the standard message mode. Alternatively, with probability (1-$R_{control}$) the travel photon is transmitted through Alice's mirror which starts the control mode. Due to a lack of detectors the control measurement setup at Alice was not completed yet. If this is done certain eavesdropping attacks that are addressed in section 4 can be carried out experimentally in addition to the here discussed encoding aspects.

In a message run the travel photon is sent through a Pockels cell with half-wave voltage of 233 V to encode the bit value by deterministically switching between the $|\psi^+\rangle$ and $|\psi^-\rangle$ state. The optical axis of the resulting half-wave plate is aligned parallel to the vertical polarization direction. The Pockels cell's switching time is limited by the rise time of the voltage of the driver. The rise time of the driver is shorter than 15 ns which is not a limiting parameter at our current transfer rates in the kHz range (see section 4). After the Pockels cell the photon is sent back from Alice to Bob (pong) where a special Bell state analysis takes place [26]. It has some added unique features that allow both for the Bell state analysis in message mode and single photon polarization characterization in control mode without active intervention.

In case of the message mode the travel and home photon enter the BS simultaneously and interfere. The resulting signature for the $|\psi^+\rangle$ state is two simultaneous detections in $D_{1,B}$ and $D_{2,B}$. The signature for $|\psi^-\rangle$ is a detection in each detector $D_{1,B}$ and $D_{2,B}$ but with one temporally delayed because of the longer propagation distance between BS and PBS for one of the photons as considered in detail in the next section. Thus, for the two different Bell states we can expect unique time stamp signatures with one detection event in each detector at Bob.

With a given probability (1-$R_{control}$) the travel photon is transmitted through Alice's mirror and a control run starts. In this case the polarization of the single travel photon gets characterized by Alice in the z-basis. At Bob automatically the polarization of the single home photon gets characterized in the z-basis as well. The home photon then enters the Bell state analyzer and propagates along the direct or detour path. Because of the half wave plate (HWP) a vertically polarized photon is always detected in $D_{1,B}$ and a horizontally polarized photon always in $D_{2,B}$ respectively. By publicly comparing their results of such control run Alice and Bob have the chance to detect an eavesdropper with a probability of 50 % per control run [1]. Unlike the original protocol, in our implementation the entire control mode sequence happens and runs automatically without any additional active switching and without the need for Alice to announce it publicly in advance. The decision for a control or a message run is totally random with a probability determined by the mirror reflectivity R$_{control}$ at Alice. During control mode Alice and Bob will get an additional indication for an eavesdropper along the lines of Wojcik's attack [17] if Alice communicates her measurement outcome not until Bob measures his home photon and both check for one and only one detection event on each side [16]. At present the control mode has not been fully implemented yet.

## 3    Photon interference at Bobs beam splitter

Within the operation of the setup presented in the last section the crucial point is the interference of both photons at the beam splitter which goes back to the Hong-Ou-Mandel interference [30]. The interference at the beam splitter will enable Bob to distinguish the two Bell states $|\psi^+\rangle$ and $|\psi^-\rangle$ as pointed out below.

To test this expected interference we varied the length of the home photon's path via a translation stage very similar to the dense coding experiment in [27] and measured the rate of conincidences between the detectors D$_{1,B}$ and D$_{2,B}$ at equal detection times and delayed detection times. This delay means the specific delay time between the two detectors D$_{1,B}$ and D$_{2,B}$ defined by the photons extra propagation time along the detour between BS and PBS which passes the half wave plate (HWP, see Fig. 1). This path via the half wave plate has a length of 1.74 m whereas the other output of BS that directly enters the polarizing beam splitter (PBS) is just 50 mm long. The difference of 1.69 m between these two beam splitter output paths amounts to a relative time delay of 5.7 ns.

Because of this extra length in one of the two beam splitter outputs the $|\psi^-\rangle$- state is expected to yield a dip at stage position 0 µm for the coincidences at equal detection times. And the $|\psi^+\rangle$-state is expected to yield a maximum for equal detection times since both photons travel along the same output path of the BS with necessarily equal flight times. The coincidences for detections delayed by the extra 5.7 ns of the longer BS-output path should produce the mirrored interference coincidence signature, meaning they yield a dip for the $|\psi^+\rangle$- state and a maximum for the $|\psi^-\rangle$- state. Thus, for both Bell states well distinguishable signatures are expected.

The accuracy of the detection of the different entangled states at Bob relies on the time accuracy of the detection evaluation of Bob's detectors. The simultaneous and delayed coincidences are evaluated using a multichannel interface (Becker&Hickl SPC 134) in time tag mode operated by a PC. To increase the robustness of the entire scheme a trigger signal provided by the pump pulse of the SPDC is used. This trigger signal marks the time slot when photons are to be expected and serves as a clock for the single photon detectors. The photons have to arrive within a common slot of ±1.5 ns or slots separated by 5.7 ns ±1.5 ns to be designated as $|\psi^+\rangle$ or $|\psi^-\rangle$-state respectively.

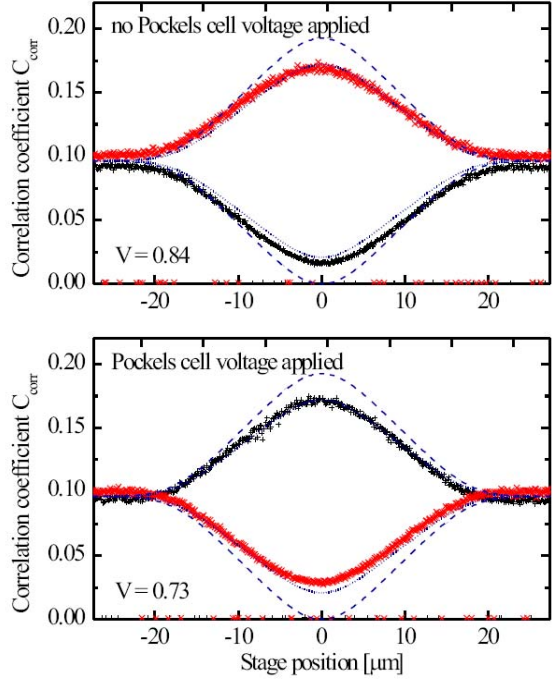

Fig. 2: Measured photon interference at beam splitter BS with Pockels-cell without applied voltage (top) and with applied voltage (bottom). The red crossed curves show the coincidences for equal detection times the black dotted curves mark the coincidence of 5.7 ns delayed detection times. The solid lines show the theoretically expected signals, taking into account a mixture share of the entangled states of 7 % and the asymmetry of the beam splitter. The dashed curves show the signals to be expected with perfect beam splitter and perfect entangled state.

For the QKD-operation the entangled photon source is adjusted to produce $|\psi^+\rangle$-states. The Pockels cell with half-wave-voltage applied should convert them to $|\psi^-\rangle$-states. With and without applied half wave voltage the coincidences with equal detection times and the coincidences delayed by 5.7 ns are evaluated. Fig. 2 shows these four measured coincidence signals as a function of the translation stage position (see Fig. 1.). As these measurements demonstrate the two Bell states are easily distinguishable at the stage position 0 µm (for equal path length of home and travel photon) and can be switched via the Pockels cell as expected. No voltage applied to the Pockels cell yields the expected signature for the $|\psi^+\rangle$-state. Half wave voltage applied to the Pockels cell yields the expected signature for the $|\psi^-\rangle$-state. The doubled width of these signals divided by c corresponds roughly to the coherence length of the biphoton D·L [29] of 240 fs (D = 2.43·10$^{-10}$ s/m the dispersion parameter of BBO and L=1mm the length of the BBO).

The contrast V = (max-min)/(max+min) between the two coincidence signals without applied Pockels cell voltage is $V^+$ = 0.84 and for the case of applied Pockels cell voltage it is $V^-$ = 0.73. The reasons for this limited contrast are the asymmetry of our beam splitter (BS) and the non perfect source of the polarization entangled photons. The BS has an asymmetric ratio of transmission and reflectance of 0.37/0.57 with losses of 6 %. The contrast reduction from $V^+$ to $V^-$ results from a lack of alignment of the detour path behind the beam splitter. The coincidences with correlation coefficient zero stem from synchronization problems in the coincidence detection. The calculated functions in Fig. 2 (solid lines) is the numerically evaluated interference of biphoton wave function [29]. They take into account the asymmetry of the beam splitter used, the bandwidth of the pump beam, the detection bandwidth of 28 nm and an estimated mixture part in the polarization entangled state of about 7 %. The small deviations of the calculated coincidence signal from the measured one result from slight differences between the reflectivities for vertical and horizontal polarization and alignment uncertainties that were not taken into account for the calculation. The mixture fraction in the entangled state and the asymmetric beam splitter should lead to an increased error rate (see section 4). Fig. 2 also shows the curves (dashed line) expected for a perfect beamsplitter and perfect entangled photon state. In this case a contrast of 1 can be restored.

## 4     Demonstration of the encoding and decoding and aspects for secure transmission

To demonstrate the deterministic information transmission we generated a random 10,000 bit long binary key. The key was transmitted from Alice to Bob in the above explained way. Due to the statistical nature of our SPDC-entangled-photon source in this prototype-implementation many pump pulses have to be used to obtain one entangled photon pair on average per encoding cycle. To keep the need for parallel classical communication and/or for key sifting as low as possible we decided to implement the information transmission with a fixed number of pump pulses per bit. After a fixed number of pump pulses the next bit was encoded by Alice. In cases of more than one detected coincidence (see below) the bit value was determined at Bob by the signature of the majority of detected coincidences within one pump pulse block. If there was an equal number of coincidences of the signatures for $|\psi^+\rangle$ - states and for $|\psi^-\rangle$ - states (reasons for this see below) the bit value was randomly set by Bob with a 50 % chance for the right choice.

Within this framework we tested the scheme with different pump pulse numbers per bit. Two of them we want to present here. First, only one detected coincidence at Bob on average per transmitted bit was used. 2500 pump pulses had to be used to yield one coincidence on average. Due to the statistic nature of our SPDC source this resulted in 40.9 % of the cases where there was no coincidence detected. In 34.4 % there was 1 coincidence, in 17,6 % 2, in 5.8 % 3, in 1.1 % 2 and in 0.2 % there were 5 coincidences detected. In a test sequence of 1000 bits 50.4 % of the bits were detected correctly (category I). For 40.9 % no bit could be evaluated due to missing detector clicks (category II). For 4.2 % of the bits a bit value could not be evaluated since there was an equal number of coincidence detections with equal numbers specific for $|\psi^+\rangle$ - states and for $|\psi^-\rangle$ - states (category III), and only 4.5 % of the bits were detected strictly wrong (category IV). In such a transmission post-processing would be necessary to communicate between Alice and Bob which bits were decoded and hence which bits could be used for the key. In this case the quantum bit error rate (QBER) would be 4.5/(50.4 + 4.5) = 8.2 % after post-processing. As QBER we define the number of detected bits at Bob where the value deviates from the encoded bit-value divided by the total number of bits used for the key.

The reasons for the number of missing and non-specific clicks are as follows:
- The statistics of the entangled-photon-source and the detection efficiencies smaller than

100 % leads to some 2500 pulse-sets in which no coincidence is detected. This dominates the category II events.
- The non-perfect entangled photon state and the asymmetric beam splitter cause the non specific detector clicks of category III and also the share of the strictly wrong detected bits (category IV).
- The lack of synchronization between the different detector channels leads to a share of 2 % non specific or missing detector clicks in category II and III.
- The double pair emission probability of 4 % can lead to non-specific signatures of category III and wrong detected bit of category IV.

Due to its spontaneous pair emission process the SPDC-source is a suboptimal choice for this protocol and should be replaced as soon as deterministic photon pair emitters which emit pairs on demand are available. In this case a coding device can be used which is only active when a photon passes by. With our SPDC-source the transmission is less efficient but is expected to be still secure as is discussed later in this section.

Second for the full proof of principle demonstration of a message transmission we used 20,000 pulses per bit to transmit a key of 10,000 bit length. At our single photon counting rate of $1.2 \cdot 10^5$/s this yields 7 detected photon pairs to transmit a single bit. After generation and transmission we used this key as a one-time pad to securely submit a message, a 10,000 bit large logo of the University of Potsdam, via a public channel from Alice to Bob. For the encoding procedure at Alice and decoding at Bob a bitwise XOR-operation was applied on the logo and the key.

In this case the QBER is 3.8 % without post-processing. The strictly wrong detected bits in category IV are 0.7 % only and the share of bits in category II and III is 4.1%. In category III half of the bits can be evaluated with correct bit value (see above). Thus, the QBER of 3.8 % is smaller than 0.7 % + 4.1 %. The error rate is not dominated by the statistics of the source anymore but by the non-perfect synchronization between the different channels of the detector readout interface. This problem is not solved yet but is solely an electronic problem. If these synchronization errors are excluded the error rate will be reduced to 1.8 %.

The resulting key transmission rate of 4250 bits/s follows from the repetition rate of the 85 MHz of the pump pulses and the block of 20,000 pulses used per transmitted bit. This relatively high transmission rate is achieved due to the deterministic character of the encoding. By the following measures the transmission rate could further be improved. The category III and IV events can quite easily be reduced by a more symmetric beam splitter compared to the one we used. A lower mixture share in the EPR-source close to 1 % is feasible and would further reduce these events. Moreover, detector readout interfaces with a better synchronization between the channels are commercially available now. They would almost eliminate the thereby caused events in category II and III. In case of the availability of a deterministic photon pair source that emits single entangled photon pairs on demand the encoding operation would be performed only when the travel photon is sent through the encoding device. Still, at a limited detector efficiency more than one pair per bit has to be sent. The rate of these deterministic emitters is not related to the probability to emit more than one pair at once. High repetition frequencies can be realized together with the guaranteed single pair emission on demand. Thereby the waiting time for coincidence detection can be strongly reduced and transmission speed can be enhanced in perspective with these future-deterministic on demand devices.

Finally, we would like to discuss single aspects of the security of our implementation by considering individual attacks. We are aware that this is not a complete investigation including collective or coherent attacks as defined in [31]. Different from single photon QKD in the ping-pong-coding scheme the travel photon is not an information carrier as pointed out above. There is no chance to extract any information if a photon is intercepted and resent. Thus, transmitting more than one photon per bit like in our mode 2 will reduce the transmission rate but is not necessarily a security risk if these photons are not sent together but one after each other.

However, due to the ping-pong characteristic and the encoding operation one has to deal with Trojan horse attacks. Eve can perform these Trojan horse attacks if she emits her own photons in a suitable Basis (x-Basis) through Alice's encoding device. From the comparison of the photons input and output polarization she could deduce the encoding operation and hence the bit value. But each time she performs such an attack she risks to get detected in a control run. To ensure that such Trojan horse attacks have to use the same mode as the ordinary transmission and control detection the encoding device has to be equipped with a mono mode spatial filter and a spectral filter with the acceptance bandwidth of the detection in the control unit.

A Wojcik attack with more than one Trojan horse photon will be detected during the control run if Alice communicates her measurement outcome not until Bob measures his home photon and both check if there is exactly one detected photon on either side [16]. Due to the low double photon emission probability of our source any double click occurrence more frequent than to be expected is a strong indication of an eavesdropper.

A specific modification of the protocol can strongly reduce the category II events but will slow down the communication distinctly. Category II events can strongly be reduced if a new bit is encoded not after a certain number of pulses but after a certain number of detected coincidences, in the optimum case after only one detected coincidence. In this case no post-processing would be necessary. Instead, a confirmation signal has to be sent from Bob to Alice. The possibility for Trojan horse attacks would be reduced, too.

In summary, we demonstrated the first implementation of the information encoding following the ping-pong coding protocol as a quantum key distribution protocol. The key is encoded in the polarization entangled states of photon pairs. Only one photon of the pair travels from Bob to Alice and back to Bob. The practicability of the implemented lab prototype scheme is currently evaluated. Replacement of the Bell state analysis including the interference at the beam splitter might be desirable and will be examined although the propagation length difference between home and travel photon is not critical in sub wavelength range. Currently a key transmission rate of 4250 bit/s is achieved. The quantum bit error rate of 3.8 % can be reduced by a better synchronization of the detector read out. The scheme can be further improved in steps by a symmetric beam splitter, a photon pair source with a negligible share of a mixture or the modification of the implementation for a deterministic entangled photon pair source.